\documentclass[preprint]{JHEP3}
\usepackage{epsfig}
\usepackage{graphicx}
\usepackage{wrapfig}
\usepackage{boxedminipage}
\usepackage{setspace}
\usepackage{amsmath}
\usepackage{yfonts}
\usepackage{subfigure}
\usepackage{dcolumn}
\usepackage{bm}
\def\be{\begin{equation}}\def\ba{\begin{eqnarray}}
\def\ee{\end{equation}}\def\ea{\end{eqnarray}}
\def\ben{\begin{enumerate}}\def\bitem{\begin{itemize}}
\def\een{\end{enumerate}}\def\eitem{\end{itemize}}
\def\no{\nonumber\\}

\newcommand{\e}{{\mbox{e}}}

\def\roughly#1{\mathrel{\raise.3ex\hbox{$#1$\kern-.75em%
\lower1ex\hbox{$\sim$}}}}

\def\A0{A_0}
\def\bq{\begin{equation}}
\def\eq{\end{equation}}

\def\K0{K^0}

\newcommand{\qn}{\textswab{q}}
\newcommand{\wn}{\textswab{w}}
\newcommand{\<}{\langle}
\renewcommand{\>}{\rangle}

\newcommand{\q}{\bm{q}}

\newcommand{\stru}{\rule[-.2in]{0in}{.2in}}

\title{ Quark number Susceptibility and  Phase Transition in hQCD Models }

\author{ Kwanghyun Jo \\
 Department of Physics, Hanyang University, Seoul 133-791, Korea}

 \author{ Youngman Kim \\
 School of Physics, Korea Institute for Advanced Study,
Seoul 130-722, Korea}

 \author{ Hyun Kyu Lee and Sang-Jin Sin \\
 Department of Physics, Hanyang University, Seoul 133-791, Korea}

\received{}
\revised{}
\accepted{}

\preprint{\hepph{}}
\abstract{
We study the quark number susceptibility, an indicator of QCD phase transition, in the hard wall and soft wall models of hQCD.
We find that the  susceptibilities in both models  are the same, jumping up at the deconfinement phase transition temperature.
We also find that the diffusion constant in the soft wall model is enhanced  compared to the one in the hard wall model.}


\begin{document}

\section{Introduction}
There has been much interest in applying the idea of AdS/CFT\cite{adscft}
 in strong interaction.
  After initial set up for N=4 SYM theory,
 confining theories were treated with IR cut off at the
 AdS space\cite{polchinski} and
 quark flavors \cite{karch} were   introduced
 by adding extra probe branes.
  More phenomenological models were also suggested
  to construct a holographic model dual to
  QCD~\cite{Brodsky,SS056,EKSS,PR}.
The finite temperature version of such model were suggested in
\cite{witten2,GY,Aharony2}.  For the purose of the Regge trajectory, quadratic dilaton  background was introduced Ref.~\cite{KKSS}  whose
 role is   to prevent string going into the deep inside the IR region of AdS space by a potential barrier and as a consequence the particle spectrum rise more slowly compared with hard-wall cutoff.   Remarkably such a dilaton-induced potential gives exactly the linear trajectory
 of the meson spectrum.
In \cite{Shuryak06}, it was argued that such a dilatonic potential
  could be motivated  by instanton effects.

The quark number susceptibility  $\chi_q$, which measures the response of QCD to a change in the chemical potential, was proposed  as a probe of the QCD chiral/deconfinement  phase transition at zero chemical potential~\cite{GLTRS, Mc87}.
It is  one of the  thermodynamic  observables  that can reveal a character of chiral phase transition.
 The lattice QCD calculation ~\cite{GLTRS} showed the enhancement of the susceptibility around $T_c$ by a factor 4 or 5.
Since then, various
 model studies~\cite{Kunihiro:91, CMT, HKRS} and lattice simulations
~\cite{GLTRS88, GPS, Allton, GG05, EKR, L07}  have been performed to
 calculate the susceptibility.

In this work, we calculate the quark number susceptibility and
study QCD chiral/deconfinement phase transition in
 holographic QCD models~\cite{EKSS, PR, GY, KKSS}. In the
 models adopted in the present work, we implicitly assume that chiral
 symmetry restoration and deconfinement take place at the same critical
 temperature $T_c$.

In a gravity dual of QCD-like model, the confinement to
 de-confinement phase transition is described by the  Hawking-Page
 transition (HPT). At low temperature, thermal AdS dominates the partition function, while
at high temperature, AdS-black hole geometry  dominates. This  was
 first discovered in the finite volume boundary case in
 \cite{witten2},
 and more recently it is shown in \cite{Herzog} that the same phenomena happen  also for infinite
boundary volume, if there is a finite scale  along the fifth direction.
In our work, both hard wall~\cite{EKSS} and soft wall~\cite{KKSS} models
 are considered, where we have a definite IR scale,
 so that we are dealing with theories with deconfinement phase transition.

In the presence of the AdS black hole, the most physical boundary condition
is the infalling one.
In the hard wall model~\cite{EKSS, PR}
 that the infalling boundary condition in the zero frequency
 and momentum limit can be understood
 as a conformally invariant Dirichlet condition.
In the the soft wall model~\cite{KKSS}, we calculate the quark number susceptibility and find that it is the same with the one obtained in the hard wall model, while diffusion constant of the soft wall model is enhanced   compared to the hard wall model.

The rest of the paper goes as follows.
In section 2, we  briefly summarize the
holographic QCD models adopted in the present work and discuss the chiral symmetry restoration
in the models.
In section 3, we calculate the quark number susceptibility
in the AdS black hole background, adopting the infalling boundary condition, in the hard wall and soft wall models.
Section 4 gives summary.
In Appendix A, we re-evaluate the susceptibility with the Dirichlet boundary condition
followed by the implication of the Hawking-Page  transition~\cite{Herzog}
\section{ Holographic QCD and chiral symmetry at finite temperature}
\subsection{ Chiral symmetry in Hard wall model:}
The action of the holographic QCD model suggested in
\cite{EKSS, PR, GY}  is given
by
\ba
&&{\rm S}_{\rm I} =\int d^4x dz \sqrt{g}{\cal L}_5\, ,\no
&&\,\,\,\, {\cal L}_5={\rm Tr} \biggl[
-\frac{1}{4 g_5^2}(L_{MN}L^{MN} +R_{MN}R^{MN} )+|D_M\Phi|^2-M_\Phi^2|\Phi|^2
 \biggr]\, ,\label{hQCDI}
\ea
where $D_M\Phi =\partial_M\Phi -iL_M\Phi +i\Phi R_M$ and
$L_M=L_M^a \tau^a/2$
with $\tau^a$ being the Pauli matrix.
 The scalar field is defined by
$\Phi=S \e^{i\pi^a\tau^a}$ and $<S>\equiv\frac{1}{2}v(z)$,
where $S$ is a real scalar and $\pi$ is a pseudoscalar. Under ${\rm
  SU(2)}_V$, $S$ and $\pi$ transform as singlet and triplet.
In this model, the 5D AdS space is compactified such that
$0<z\leq z_m$.
\footnote{  $z_m$ is a infrared (IR)
cutoff, which is fixed by
the rho-meson mass at zero temperature: $1/z_m \simeq 320~ {\rm MeV}$,
and the value of the 5D gauge coupling
 $g_5^2$ is identified as $g_5^2=12\pi^2/N_c$ through
matching with QCD~\cite{EKSS, PR, KKSS}.}

 As in~ \cite{GY}, we work on the  5D AdS-Schwarzchild background,
which is known to describe the physics of the finite temperature in
dual 4D field theory side,
\ba
ds_5^2=\frac{1}{z^2}\biggl(f (z) dt^2-(dx^i)^2 -\frac{1}{f (z)}dz^2
\biggr),~~f (z)=1-\frac{z^4}{z_T^4},\label{bhs}
\ea
where $i=1,2,3.$
Here the temperature is defined by $T=1/(\pi z_T)$, and the Hawking-Page
transition~\cite{Herzog}  occurs at $T_c=2^{1/4}/(\pi z_m)$. For the temperature lower than $T_c$, thermal AdS dominates, and it is hard to find temperature dependence in low temperature regime, which is actually consistent with the earlier work on large N gauge theory \cite{pisarski}.

The equation of motion for $v(z)$ in the black hole background  is
\ba
\biggl[ \partial_z^2 -\frac{4-f }{zf }\partial_z +\frac{3}{z^2f }
\biggr]v(z)=0\, ,
\ea
and the solution is given  by~\cite{GY}
\ba
v(z)= M_q z~{}_2F_1
(\frac{1}{4},\frac{1}{4},\frac{1}{2},\frac{z^4}{z_T^4})
+  \Sigma_q z^3 ~{}_2F_1
(\frac{3}{4},\frac{3}{4},\frac{3}{2},\frac{z^4}{z_T^4})  .
\ea
Here $M_q$ and $\Sigma_q$ are identified with the current quark mass   and
the chiral condensate  respectively.
Note that at $z=z_T$, both terms in $v(z)$ diverges logarithmically.
This requires us to set  both  of them  to be zero,
\be
M_q=0, \quad \quad   \Sigma_q =0.\ee
 The latter condition means
 that the chiral symmetry is restored,  and deconfinement   and the chiral phase transition take place at the same temperature.
It is interesting to observe that the mass term is also forbidden in this phase.  
This is consistent with the fact that the chiral symmetry  forbids fermion mass term. \footnote{Notice  that we never discussed any fermions in this formalism, and hence we never defined any chiral symmetry in this theory explicitly.}  In reality  the chiral symmetry is partially broken in low temperature and also partially restored in the high temperature due to the current quark mass.
In this sense, the hard wall model respects the chiral symmetry
more than the reality.
\footnote{In D3-D7 system, $\Sigma_q, M_q$ are not necessarily zero
 even for deconfined phase.
}

\subsection{ Chiral symmetry in Soft wall model} In \cite{KKSS}, dilaton background was introduced for the Regge behavior of the
spectrum.
\ba
{\rm S}_{\rm II} =\int d^4x dz\e^{-\Phi}{\cal L}_5  \, ,\label{hQCDII}
\ea
where $\Phi=cz^2$.
Here the role of the  hard-wall IR cutoff $z_m$
is replaced by a dilaton-induced potential.
The equation of motion for $v(z)$  is given by
\ba
\biggl[ \partial_z^2 -(2cz+ \frac{4-f }{zf })\partial_z
 +\frac{3}{z^2f }
\biggr]v(z)=0 \, .\label{EoMvz}
\ea
At zero temperature \cite{KKSS}, where  $f =1$, one of the two linearly independent
solutions of Eq. (\ref{EoMvz})
 diverges as $z\rightarrow \infty$, and
so we have to discard this solution.
Then chiral condensate is simply proportional to $M_q$~\cite{KKSS}.
Now we  consider a finite temperature case.
Near $z_T$, $c$ dependent term  in Eq. (\ref{EoMvz}) is negligible
and there is no difference between hard and soft wall model near the horizon. So we can draw the same conclusion of the complete chiral symmetry restoration.

\section{Quark number susceptibility\label{sec.3}}
In this section, we calculate the quark number
susceptibility   at high temperature in deconfinement phase.
 The relevant background is the AdS black hole,
\be
ds^2 = \frac{(\pi T)^2}{u} \bigg(f(u)dt^2 - (dx^i)^2 \bigg) - \frac{1}{4u^2f(u)} du^2\label{AdSBHu}
\ee
where u=$(z/z_T)^2$, f(u) = 1-$u^2$ and T = $1/\pi z_T$.

The quark number susceptibility was proposed  as a probe of the
QCD chiral phase transition at zero chemical potential~\cite{GLTRS,Mc87}.
\footnote{In this work, we will not distinguish flavor singlet and non-singlet susceptibilities, $\chi_S$
and $\chi_{NS}$ respectively.  Lattice QCD studies showed
 that $\chi_S\simeq \chi_{NS}$~\cite{GLTRS, GLTRS88,GPS, EKR}, which is because
 the mixing between isospin singlet and triplet vector mesons, $\rho$ and $\omega$, is
 tiny \cite{Kunihiro:91}.}
 \be
 \chi_q=\frac{\partial n_q}{\partial\mu_q}.
 \ee
The quark number susceptibility
can be written in terms of the retarded Green's function. Here we follow the procedure given in \cite{Kunihiro:91}.
 First we write
\be
\chi_q ( T, \mu) =\beta \int d^3x G_{00} ( 0, x ),
\ee
where the vector correlator  at finite temperature   is defined by
\be \label{wightmanfunction}
G^{\mu\nu}(p ;T) \delta_{ab}
=
\int_0^{1/T} d \tau \int d^3\vec{x}
e^{-ip\cdot{x} }
\left\langle
  J_{ a}^\mu(\tau,\vec{x}) J_{  b}^\nu(0,\vec{0})
\right\rangle_\beta \, .\ee
After the Fourier transformation,
we obtain
\be
\chi_q  ( T, \mu)=\lim_{k \rightarrow 0} \int^{\infty}_{-\infty} \frac{d \omega}{2 \pi}
 G_{00} (\omega, k)
\ee
By the fluctuation-dissipation theorem, we have
\be
G _{00}(\omega, k) = -\frac{2}{1-{\mathrm {e}}^{-w/T}} {\mathrm{Im}}G^R_{00}(\omega,k)
\ee
where $G^R_{00}(\omega,k)$ is retarded Greens function of $j_\mu, j_\nu$.
Then we arrive at
\ba
\chi_q ( T, \mu)
 =  -\lim_{k \rightarrow 0} \int^{\infty}_{-\infty} \frac{d \omega}{2 \pi}
 \frac{2}{1-{\mathrm {e}}^{-w/T}} {\mathrm{Im}}G^R_{00}(\omega,k)
\ea
The $\mathrm{Im} G_{00}^R(\omega,k)$ is related to the real part of Green's function
through the Kramers-Kronig dispersion relations,
\be
\mathrm{Re} G_{00}^R(\omega,k) =
\mathcal{P} \int^{\infty}_{-\infty} \frac{d \omega'}{\pi}
\frac{\mathrm{Im} G^R_{00}(\omega,k)}{w'- w}\, .
\ee
One can easily see that $\chi_q$ is written  in the following form,
\be \label{defofsusc}
\chi_q(T,\mu_q) = -\lim_{k \rightarrow 0} \mathrm{Re} G_{00}^R(0,k)\, .
\ee
What is the physics we want to see?
 The lattice QCD calculations ~\cite{GLTRS, GLTRS88, GPS, EKR} showed the enhancement of $\chi_q$ around $T_c$:
\ba
R_\chi\equiv \frac{\chi_q (T_c+\epsilon)}{\chi_q (T_c-\epsilon)}=4 \sim 5\, .\label{Rv}
\ea
The enhancement in $R_\chi$ may be understood roughly as follows~
\cite{Mc87}. At low temperature, in confined phase $\chi_q$ will pick
up the Boltzmann factor $\e^{-M_N/T}$, where $M_N$ is a typical hadron
mass scale $\sim 1~{\rm GeV}$, while at high temperature the factor will be given
in terms of a quark mass  $\e^{-M_q/T}$, and therefore there could be
 some enhancement.
In Ref.~\cite{Kunihiro:91}, it is shown that the enhancement in $\chi_q$
may be due to the vanishing or a sudden decrease of the interactions
between quarks in the vector channel.

In the holographic QCD models, due to the HPT~\cite{Herzog},
 the quark number susceptibility  is described by the AdS black hole background at high  temperature   and
by the thermal AdS at low
temperature. We will patch them together.  We
 calculate the  susceptibility both in the hard wall model~  and in the soft wall model and describe how  
 it changes under the phase transition.

  \subsection{ Hard wall model}

The temperature dependence of confinement phase can not be extracted from the AdS black hole
due to the Hawking-Page transition.
That is, thermal  AdS replaces the AdS black hole  at low temperature.
Therefore we need to consider two phase separately.

\subsubsection{Confined phase\label{311}}

We take the  thermal AdS with hard wall at $z=z_m$ as the dual gravity background.
{}From the quadratic part of the 5D action
Eq. (\ref{hQCDI}), we obtain the equation of motion for the time
 component vector meson
\ba
\biggl[\partial_z^2-\frac{1}{z}\partial_z +  \vec{q}^2 \biggr]V_0
(z,\vec q)=0.
\label{EoMV}\,
\ea
Note that  in the above equation, the $q_0\to 0$ limit is already taken, following the definition given in
 Eq. (\ref{defofsusc}). With a finite $q_0$ the equation will couple to other components.
We solve Eq. (\ref{EoMV}) in a
limit where ${\vec q}\to 0$.
The solution is $V_0=a_1-a_2z^2 \, .$
According to the AdS/CFT dictionary, $V_0$ is dual to the number density $j^0(x)$  in boundary, hence
we can identify $a_1$ as the chemical potential $\mu$
 and the $a_2$  as the charge density $Q$.
Therefore
\ba
V_0=\mu-Q(z/z_m)^2 \, .
\ea
Then the quark number susceptibility is given by
\ba
\chi_q(T)&=&- \left [\frac{1}{g_5^2}\frac{\partial_z V_0}{z} \right]_{z=0}
= \frac{2Q/z_m^2}{g_5^2}\, .
 \label{chi1}
\ea
For the black hole background, we have to request $V_0=0$ at the horizon for the regularity of the vector field, making $\mu$ proportional to $Q$. However, for the thermal AdS  there is no such requirement to be imposed. Therefore, even when a charge is zero,
 one can still have a finite chemical potential in confining case.
However, in the limit where the chemical potential is zero, the charge density must be zero, otherwise we would need work to bring a particle into the system, requesting a finite chemical potential. That is,  in the limit  of zero chemical potential we should have zero charge and hence
\be
\chi_q=0,
\ee
in the confined phase.
  \subsubsection{Deconfined phase}
 Here we calculate the quark number susceptibility in deconfined phase on the metric~(\ref{AdSBHu}).
Note that the role of the IR cutoff $z_m$ is none  in the deconfined phase. The action of the gauge field, which is dual to the 4D quark current $j_\mu=\bar q(t,\vec x) \gamma_\mu q(t,\bar x)$,
is
\be \label{actionHard}
S = -\int d^5 x \sqrt{g} \frac{1}{4g_5^2} F_{MN}F^{MN}\, ,
\ee
where $1/g_5^2=N_cN_f/(2\pi)^2$ coming from the D3/D7 model \cite{sin}.
In the presence of the AdS black hole, the most natural boundary condition  is the infalling boundary condition since the black hole can only absorb classically. For the static case, we may consider the Dirichlet and Neumann boundary conditions.
To impose the infalling boundary condition, we solve the problem at small but non-zero frequency and momentum and then take them to be
 zero.    This is the problem considered in the literature on hydrodynamics \cite{hydro}.
 Notice that
in terms of the re-scaled coordinate $u=(z/z_T)^2$,
the  momentum $k=(\omega,0,0,q)$ enters in the action only
in the combination of  $  \wn = \frac\omega{2\pi T}, \quad \qn = \frac q{2\pi T}$.
The relevant equations of motion for the vector fields  in the
$A_u=0 $ gauge are
\begin{eqnarray}
  &&  \wn A_0' + \qn f \, A_3' = 0 \stru \,,\quad  A_0'' - \frac1{uf}\left( \qn^2 A_0 + \wn \qn A_3 \right) =0\,,
  \stru \label{eq2}\\
  &&  A_3'' + {f'\over f} A_3'  + \frac1{uf^2}
 \left(\wn^2 A_3 + \wn \qn A_0 \right) = 0\,,
  \stru \label{eq3}
\end{eqnarray}
where $'$ means $\partial_u$.
Out of the coupled equations we can eliminate $A_3$ to get
a second order differential equation for $A'_0:=\Psi$.
\begin{equation}
 \Psi'' + \frac{(uf)'}{uf} \Psi' +
  {\wn^2 - \qn^2 f(u)\over  u f^2}\Psi=0\,,
\label{eq6}
\end{equation}
After taking out the near  horizon behavior $\sim (1-u)^{- i\wn/2}$ of
  infalling wave, one can extract the small frequency behavior of the
  residual part using Mathematica.  The result is
\be
\Psi = (1-u)^{- i\wn/2} \cdot \frac{\qn^2 A_0^0 + \wn \qn A_3^0}{(i\wn-\qn^2)} \Bigl(1+\frac{i\wn}{2} \ln \frac{2u^2}{1+u}-\qn^2 \ln \frac{2u}{1+u} + \cdots \Bigr),
\ee
where $A_\mu^0$ is the boundary value of $A_\mu$.
With the prescription for the retarded Green function
\begin{equation}\label{GR-prescr}
  G_{\mu\nu}^R(k) =  \frac{\delta^2 S}{\delta A_\mu^0(-k)\delta A_\nu^0(k)},
\end{equation}
with
\be
S=\frac{\pi^2 T^2}{g_5^2}\int d^4k \Bigl[\Psi(-k)\Psi(k) +\cdots\Bigr]_{u=0}.
\ee
one can get the retarded Green function
\begin{equation}
\begin{split}
  G^R_{\mu\nu}(\omega,\q) &= -i\!\int\!d^4x\, e^{-iq\cdot x}\,
  \theta(t) \<[J_\mu(x),\, J_\nu(0)]\>
\end{split}
\end{equation}
in small frequency;
\ba
G^R_{0 0} &=& \frac{ 2\pi^2 T^2}{g_5^2}{ \qn^2  \;   \over   (i\wn - \qn^2 +i\wn^2\ln 2 /2)}\Bigl(1+ \ln2\, \Bigl(\frac i2\wn -\qn^2\Bigr)\Bigr)
\ea
Finally, by using eq. (\ref{defofsusc}), we get the susceptibility
\be
\chi_q = \frac{2\pi^2 T^2}{g_5^2}=\frac{N_cN_f}{2} T^2\, . \label{VSUShw}
\ee
$\chi_q/T^2$ is a constant which is consistent with high temperature behavior of lattice result ~\cite{GLTRS,GLTRS88, GPS, Allton, GG05, EKR, L07}.
Our results show that   $\chi_q(T)/T^2$
jumps from zero, corresponding to the thermal AdS phase,  to a
constant, AdS black hole phase, with increasing the temperature.
We plot $\chi_q$ as a function of temperature.

We close this section with an interesting
observation. If we impose the Dirichlet condition at the horizon, $V_0 (z_T)=h$, we get
\ba
\chi_q(T)=
 \frac{2\pi^2}{g_5^2} (1-h) T^2 \, ,\label{chi2}
\ea
Notice that for the conformally invariant choice
$h=0$, Dirichlet condition gives the identical result to the
infalling BC obtained above.

\subsection{ Soft wall model}

Now we consider the soft wall model.
The action is given by
\be \label{actionsoft}
S =- \int d^5 x \sqrt{g} \mathrm{e}^{-\Phi} \frac{1}{4g_5^2} F_{MN}F^{MN}  , \quad \Phi = \frac{c}{(\pi T)^2} u .
\ee
In this model, there is also the Hawking-Page transition~\cite{Herzog}. 
The equation of motion for $V_0$ in the static-low momentum limit reads
\ba
\partial_z (\frac{1}{z}\e^{-cz^2}\partial_z V_0 )=0\, ,\label{EoMV0}
\ea
whose solution is given by
\ba
V_0&=&a\e^{cz^2} +b\, ,\no
&\simeq&c_1 +c_2 z^2\, , z\rightarrow 0\, .
\ea
Then, following the same argument given in (\ref{311}), we conclude that 
the quark number susceptibility is also zero in the soft wall model at low temperature. 

Now we consider high temperature phase.
 The relevant equations of motion are
\ba
A_t : & A_t'' - \frac{c}{(\pi T)^2} A_t' - \frac{1}{u f(u)} (\qn^2 A_t + \qn \wn A_z) =0 & \label{EoMSW1} \\
A_\alpha : & A''_\alpha + (\frac{-c}{(\pi T)2} + \frac{f'(u)}{f(u)})A'_\alpha
+ \frac{1}{uf(u)}(\frac{\wn^2}{f(u)} -\qn^2)A_\alpha =0 & \label{EoMSW2} \\
A_z : & A''_z + (\frac{-c}{(\pi T)2} + \frac{f'(u)}{f(u)})A'_z
+ \frac{1}{uf(u)^2}(\wn\qn A_t -\wn^2A_z )=0 & \label{EoMSW3} \\
A_u : & \wn A'_t + \qn f(u) A'_z = 0 & \label{EoMSW4}
\ea
From the eq.(\ref{EoMSW1}), we can express $A_z$ in terms of $A_t$
\be
A_z = \frac{u f(u)}{\qn \wn} \left(A''_t - \frac{c}{(\pi T)^2}A'_t - \frac{\qn^2}{u f(u)} A_t\right)\, .
\ee
Inserting it to eq.(\ref{EoMSW4}), we obtain
\be \label{infallsoft}
A''' + \Big( \frac{1-3 u^2}{uf} - c_T \Big)A'' + \frac{1}{uf^2} \Big(\wn^2 -f ( c_T (1-3u^2) + \qn^2 ) \Big) A' = 0
\ee
where $c_T = c/(\pi T)^2$.
Imposing the infalling boundary condition at the horizon, we take  $A'_t = (1-u)^{-i \mathbf{w}/2}$ F(u) to obtain
\ba
&& F'' + (-\frac{c}{(\pi T)^2} +\frac{1-3u^2}{uf(u)} + i \frac{\mathbf{w}}{1-u})F' \nonumber \\
&& + \bigg(-\frac{c}{(\pi T)^2} (\frac{1-3u^2}{uf(u)}) + \wn \Big(\frac{i (1+2u)}{2uf(u)} -
\frac{c}{(\pi T)^2} \frac{i(1+u)}{2f(u)} \Big) \nonumber \\
&& + \wn^2 \frac{4-u(1+u)^2}{4uf(u)^2} - \frac{\qn^2}{uf(u)}\bigg)F = 0.
\ea
In the long-wavelength, low frequency limit, F(u) can be expanded as series in $\wn, \qn^2$,
\be
F(u) = F_0 + \wn F_1(u) + \qn^2 G_1 + \cdots
\ee
After some algebra, we obtain  first a few terms
\ba
F_0 & = & B \\
F_1 & = & \frac{i B \mathrm{e}^{-c_T}}{12} \bigg( 2 (\mathrm{e}^{c_T} - \mathrm{e}^{c_T u})
+ (-12 + 5 c_T) \mathrm{e}^{c_T(1+u)} \Big(Ei(-c_T) - Ei(-c_T u) \Big) \rangle \nonumber \\
& & - 2\mathrm{e}^{c_T(2+u)} (-3 + c_T) \{ Ei(-2c_T)-Ei(-c_T(1+u)) \} \bigg)  \\
G_1 & = & B \mathrm{e}^{c_T u} \bigg( Ei(-c_T) - Ei(-c_T u) + \mathrm{e}^{c_T} \Big(Ei(-c_T(1+u)) - Ei(-2c_T) \Big) \bigg)\, ,
\ea
where Ei(x) is  defined as the principal value of $Ei(z)=-\int_{-z}^{\infty } \left.e^{-t}\right/t \, dt$.
The integration constants of $F_1$, $G_1$ are chosen by the regularity condition at u=1. B is determined from the boundary values of $A_t$ and  $A_z$ at u=0.
\ba
B = \frac{\qn^2 A_t^0 + \qn \wn A_z^0 }{i \wn (1  - \frac{5}{12} c_T) - \qn^2 }.  \ea
From these results, $A_t'(u)$ is
\be
A'_t(u) = (1-u)^{-i \mathbf{w}/2} \frac{\qn^2 A_t^0 + \qn \wn A_z^0 }{i \wn (1  - \frac{5}{12} c_T) - \qn^2 } \bigg(1+ \qn^2 X_{\qn^2} +i\frac{\wn }{12}Y_{\wn}\bigg)\, ,
\ee
where $X_{\qn^2}, Y_{\wn}$ are
\ba
X_{\qn^2} &=& \mathrm{e}^{c_T u}\bigg( Ei(-c_T)-Ei(-c_T u) -\mathrm{e}^{c_T} \Big(Ei(-2 c_T)-Ei(-c_T (1+u))\Big) \bigg) \no
Y_{\wn} &=& \mathrm{e}^{c_T u}\bigg(2 \mathrm{e}^{-c_T u}-2 \mathrm{e}^{-c_T}+ (-12+5 c_T)\Big(Ei(-c_T)-Ei(-c_T u)\Big) \no
&&-2 \mathrm{e}^{c_T} (-3+c_T)  \Big(Ei(-2 c_T)-Ei(-c_T(1+u))\Big)\bigg)\, . \ea
The real part of the retarded Green's function
\be
\mathrm{Re} G_{00}^R(k) = -\frac{2 \pi^2 T^2}{g_5^2}  \frac{q^2}{P^2 w^2+\bar{D}^2 q^4}
\left(\bar{D}^2q^2 +\bar{D}^4 q^4 X_{\qn^2}(\epsilon)-\frac{w^2}{12}\bar{D}^2 P Y_{\wn}(\epsilon) \right)\, ,
\ee
where $\bar{D}=1/2\pi T$ and $P=1-5c/12\pi^2 T^2$.
Note that the Green function has diffusion pole, and the  diffusion constant is
\be
D= \frac{1}{2 \pi T}\frac{1}{1- \frac{5c} {12\pi^2 T^2}}.
\ee
Notice that it is dressed by the factor $1/P$  due to the effect of the soft wall.
 One can easily check the positivity of the diffusion constant
 in the relevant temperature regime
 \be
P= 1-0.17\frac{T_c^2}{T^2} > 0,
\quad {\mathrm{if} ~ T>T_c}
\ee
where $T_c = 1/\pi^2 {z_c}^2$, and we used  $c {z_c}^2 = 0.42$   \cite{Herzog}.

The quark number susceptibility is obtained with eq.(\ref{defofsusc})
and we get
\be
\chi_q =  \frac{2 \pi^2 T^2}{g_5^2}\, ,\label{qSUS}
\ee
which is the same with the result of the hard wall model.

Putting together the results at low and high temperatures, we arrive at the following conclusion.
  $\chi_q$ is zero up to the phase transition temperature, and it
 jumps to a finite  value given in Eq. (\ref{qSUS}), which implies a first order phase transition between low- and high-temperature
phases. The sharp transition  might be the large $N_c$ artifact.

In hard wall case we observed that the infalling and (a specially chosen) Dirichlet boundary conditions give the same results.
One may wonder if one can arrive at the same conclusion in the soft wall model.
In Appendix we dig into this question to observe that those two boundary conditions lead to
 different susceptibilities.

\section{Summary }
We first discussed the chiral symmetry restoration in AdS/QCD models.
The AdS/QCD models respect the chiral symmetry more rigidly than
the reality in the sense that, in chiral symmetry restored
phase,  both of the chiral condensate and the mass of the quarks are zero.

Then, we calculated the quark number susceptibility in both hard wall and
soft wall models.
At low temperature, in confined phase, we showed that $\chi_q$, which is defined in the limit of
zero chemical potential, is zero.  
With the infalling boundary condition, we could uniquely determine the overall normalization
 of the susceptibility, unlike the  Dirichlet boundary condition.
 We found that the  susceptibilities in both models  are the same with the
 infalling boundary condition, and $\chi_q\sim T^2$ at high temperature, which is consistent with high-temperature
 lattice QCD observations~\cite{GLTRS88, GPS, Allton, GG05, EKR, L07}.

 In Appendix A, we considered   Dirichlet boundary condition in the soft wall model.
With the HPT, we predicted  the temperature dependence of $\chi_q$ at high
 temperature apart from the overall normalization that is fixed by an IR boundary condition.
 Our result with the HPT exhibits a similar behavior observed in
 model studies~\cite{Kunihiro:91, CMT, HKRS} and lattice simulations
~\cite{GLTRS88, GPS, Allton, GG05, EKR, L07}.

Regardless of the IR boundary conditions, our results in both models predicted a sharp jump in the quark number susceptibility,
which is an unavoidable aspect of the HPT and
could be smoothed out by including large $N_c$ corrections.

Finally, we discuss a limitation of our approach in the light of the QCD phase transition.
The nature of the QCD transition depends on the number of quark flavors and the quark mass: for pure SU(3) gauge theory,
 it is a first order, for two massless quarks, it is a second order, for two quarks with finite masses, it is a cross over,
 for three degenerate massless quarks, it is a first order, etc.
 Unlike the Polyakov loop or  chiral condensate, the quark number susceptibility is not an order parameter, and so
 in the present study we are not able to determine the order of the QCD phase transition. The susceptibility
 could serve, at best, as an indicator of the transition.

\appendix

\section{Dirichlet boundary condition in soft wall model}
Here we give analysis with Dirichlet boundary conditions in high temperature.
The equation of motion for $V_0$ in the static, zero momentum limit   is the same as eq.(\ref{EoMV0}) and the solution is still given by
 $V_0=a\e^{cz^2} +b$.
 However, with
 the Dirichlet boundary conditions $V_0(0)=1, \;\;V_0(z_T)=0$,
  the result is given by
\ba
\chi_q (T)=  \frac{2c}{g_5^2} \frac{1}{\e^{{\tilde T}_c^2/T^2}-1}\label{VSUSs}
\ea
We note here that $\chi_q ({ T}_0)\approx 1.2{\tilde T}_c^2 $, where ${  T}_0=\sqrt{c}/\pi$ is the temperature scale generated by $c$.

\begin{figure}[!ht]
\begin{center}
\subfigure[] {\includegraphics[angle=0,
width=0.4\textwidth]{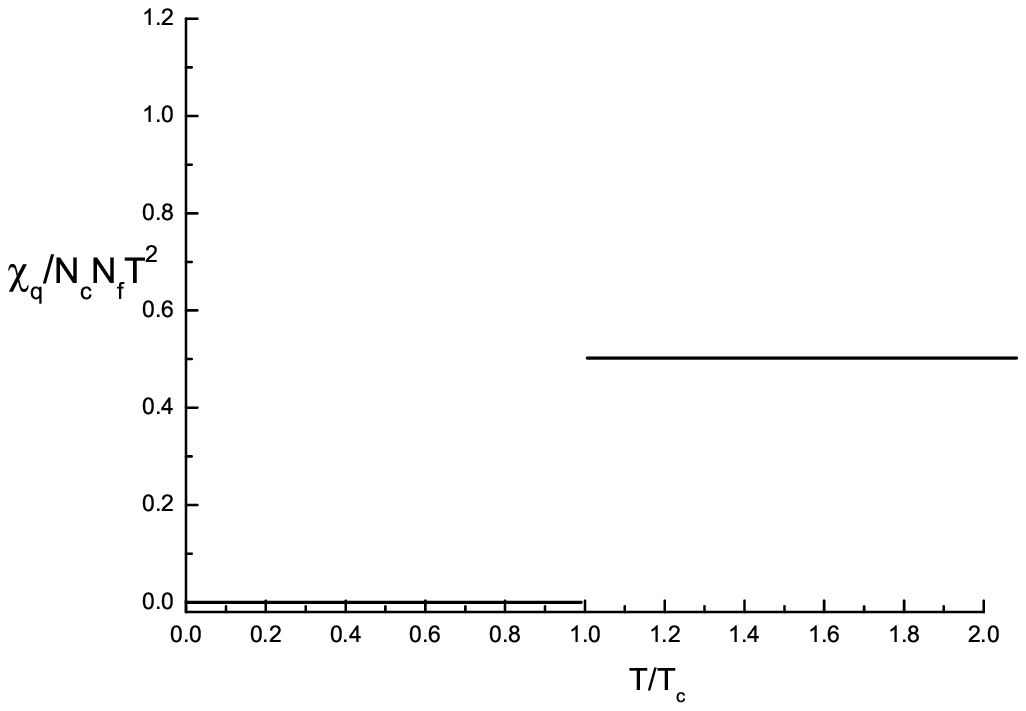} \label{hydro}}
\subfigure[] {\includegraphics[angle=0,
width=0.4\textwidth]{{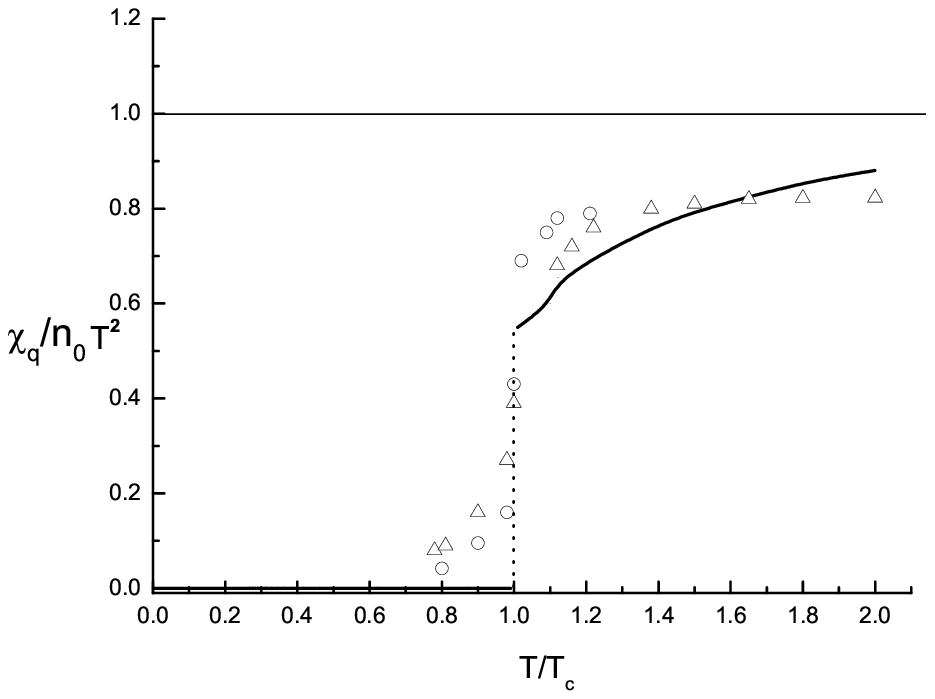}} \label{fig:sus2}}
 \caption{ (a) $\chi_q$ in hQCD with the innfalling boundary condition discussed in section 3. 
 (b) $\chi_q$  in the soft wall model with the Dirichlet boundary condition. The circles are for the
 (quenched) lattice QCD results
   as shown in Ref.~\cite{Kunihiro:91, GPS}, and the triangles are for full lattice QCD in~\cite{Allton}. }
\end{center}
\end{figure}
\section{parameters of the soft-wall models}
The masses of the vector mesons are given by
 $ m_n^2=4c(n+1)$.
If we use $m_1=770 ~{\rm MeV}$ and  $m_2=1450 ~{\rm MeV}$ to calculate
the slope of the Regge trajectory, then we obtain
$\sqrt{c}\simeq 614 ~{\rm MeV}$ and so end up with the reasonable
value of the transition temperature $T_c \simeq 195 ~{\rm MeV}$.
The value of $T_c$ was determined in
  Ref.~\cite{An0607}, $T_c=210~{\rm MeV}$, and
more recently  the relation between $T_c$ and $z_m$
  ($\sqrt{c}$) is obtained through Hawking-Page analysis in the
  holographic models used in this work, where $T_c\approx 191~{\rm
    MeV}$~\cite{Herzog}.
We note here that in \cite{An0603}, the value of $\sqrt{c}$ was
determined to be $\sim 671~{\rm MeV}$.
Finally, we relate $c$ with the QCD string tension $\sigma$.
The masses of vector towers are given, in terms of $\sigma$, by
$m_n^2=2\pi\sigma n$. From this and $ m_n^2=4c(n+1)$,  we get
$c=\frac{\pi}{2}\sigma,$ so that the dilaton factor becomes
$\e^{-\frac{\pi}{2}\sigma z^2}$. Note that the relation between $c$
and string tension was also observed in Ref.~\cite{An0604}.

\vskip 1.0cm

\noindent
{\large\bf Acknowledgments}\\
We thank Seyong Kim for useful information on lattice QCD and Ho-Ung Yee for helpful discussions.
The work of SJS  was supported by
the SRC Program of  the KOSEF through the Center for Quantum
Space-time(CQUeST) of Sogang University with grant number R11 - 2005
- 021 and also by KOSEF Grant R01-2007-000-10214-0.
The work of KJ is supported in part by the Seoul Fellowship.

\end{document}